\documentclass[conference]{IEEEtran}
\IEEEoverridecommandlockouts
\usepackage{cite}
\usepackage{amsmath,amssymb,amsfonts}
\usepackage{algorithmic}
\usepackage{graphicx}
\usepackage{tikz}
\usepackage{textcomp}
\usepackage{xcolor}
\def\BibTeX{{\rm B\kern-.05em{\sc i\kern-.025em b}\kern-.08em
    T\kern-.1667em\lower.7ex\hbox{E}\kern-.125emX}}

\newcommand\copyrighttext{%
  \footnotesize © 2023 IEEE.  Personal use of this material is permitted.  Permission from IEEE must be obtained for all other uses, in any current or future media, including reprinting/republishing this material for advertising or promotional purposes, creating new collective works, for resale or redistribution to servers or lists, or reuse of any copyrighted component of this work in other works.}
\newcommand\copyrightnotice{%
\begin{tikzpicture}[remember picture,overlay]
\node[anchor=south,yshift=10pt] at (current page.south) {\fbox{\parbox{\dimexpr\textwidth-\fboxsep-\fboxrule\relax}{\copyrighttext}}};
\end{tikzpicture}%
}

\begin{document}

\title{Pulse Compression Probing for Tracking Distribution Feeder Models\\}

\author{\IEEEauthorblockN{Nicholas Piaquadio, N. Eva Wu, Morteza Sarailoo, Jianzhuang Huang}
\IEEEauthorblockA{\textit{Department of Electrical and Computer Engineering} \\
\textit{Binghamton University}\\
Binghamton, NY USA\\
\{npiaqua1,evawu,msarail1,jhaung5\}@binghamton.edu}
}

\maketitle

\begin{abstract}
A Pulse-Compression Probing (PCP) method is applied in time-domain to identify an equivalent circuit model of a distribution network as seen from the transmission grid. A Pseudo-Random Binary Pulse Train (PRBPT) is injected as a voltage signal at the input of the feeder and processed to recover the impulse response. A transfer function and circuit model is fitted to the response, allowing the feeder to be modeled as a quasi-steady-state sinusoidal (QSSS) source behind a network. The method is verified on the IEEE 13-Node Distribution Test System, identifying a second order circuit model with less than seven cycles latency and a signal to noise ratio of 15.07 dB in the input feeder current.
\end{abstract}

\copyrightnotice
\section{Introduction}

The authors' ultimate interest lies with modernizing protective control of the evolving power grids  \cite{TSG_18} \cite{ACC_22}. The National Electric Reliability Council (NERC)'s 2021 Long-Term Reliability Assessment \cite{NERC} highlights major challenges in highly renewable grids, including demand uncertainty caused by Distributed Energy Resources (DER) on distribution feeders, as well as load weather-dependence and increased risks during extreme weather. The variability of DER and renewables raises challenges from the point of view of the transmission grid, the models for which typically use a constant power load, constant impedance/current/power (ZIP) load, or a composite load model to represent entire distribution networks.

In this paper, the transmission system is treated as a multiple-port circuit in relation to all generating units and all distribution circuits interconnected to it. Each generating unit is modeled as a quasi-steady-state-sinusoidal (QSSS) voltage source behind its internal impedance, whose phase and magnitude are modulated \cite{QSSS}, \cite{IET18} and each passive distribution circuit is modeled as a low order electric network.

Urgent needs have risen to model an active (IBR-penetrated) distribution circuit by integrating the above modeling schemes into real-time tracking of distribution feeders. We propose to apply pulse compression probing (PCP) \cite{wu_patent} for this purpose.

PCP excites a system by compressing a low magnitude pseudo-random binary pulse train in continuous-time (PRBPT) \cite{Probe_Power}, also known as PRBS in digital form \cite{PRBS_Book}, \cite{PRBS_Book2}, into a phantom impulse in a potentially high order, time-varying, and nonlinear dynamic system. At the probing output of a selected signal transfer path, a nonparametric small signal model is extracted in the form of an impulse response delayed by its own memory length \cite{NASA_Probe}. We have tested the PCP principle for health monitoring on a nonlinear aircraft model \cite{ACC11}, and a power transfer path in a High Voltage Direct Current (HVDC) transmission system model \cite{Probe_Power}. 

The main challenges in applying the PCP have to do with the probing signal design which must achieve low intrusion to the normal operation of the system being probed, high accuracy of the extracted nonparametric model for the selected signal transfer path, small time delay in tracking the time-varying behavior of the transfer path, and moderate processing to serve real-time applications. The reader is referred to \cite{Probe_Power} for an analysis on the PCP’s potential to outperform, in both accuracy and speed, the multi-sine probing implemented to identify the electromechanical modes in the Western Interconnection \cite{Hauer},\cite{WECC_Probe}.  The nonparametric modeling framework of the PCP offers simplicity, flexibility, and accuracy for protective control via monitoring and change detection over the more traditional parametric modeling framework in complex systems \cite{Gertler}, \cite{PRBS_Book2}. 


In this paper, a reduced-order equivalent model of a distribution network is obtained using only terminal measurements at the transmission substation.  PCP is implemented as a time-domain signal in phase voltage, returning a model on the time scale of cycles versus the seconds typically required for probing techniques implemented via phasor measurements \cite{survey}. This allows the impulse response of the target distribution feeder to be rapidly extracted. A QSSS model of a voltage source is placed behind the identified network, allowing near real-time dynamic tracking for distribution feeders with active sources.

The remainder of this work is organized as follows. Section II provides background on PCP and Section III develops a methodology to apply PCP in time domain, as well as fit a circuit model to the results.  Section IV implements the PCP algorithm to identify the time-domain impulse response of and an equivalent model for Phase A of the IEEE 13-bus distribution test system. Section V contains conclusions and directions for future work.
\vspace{-1mm}
\section{Review of Probing Algorithm}

This section reviews the principles and main results of pulse compression probing (PCP) following \cite{NASA_Probe}. For an unknown system the objective is to identify the system impulse response, $h(t)$. To do this, the probing signal $p(t)$ is constructed according to (\ref{probe_sig}) to form a pseudo-random binary pulse train (PRBPT), an example of which is shown in Fig. 1.  

\begin{equation}
    p(t) = \alpha \eta(t) * \sigma(t)
    \label{probe_sig}
\end{equation}

\begin{figure}[h!]
    \vspace{-3mm}
    \centering
    \includegraphics[scale=0.35]{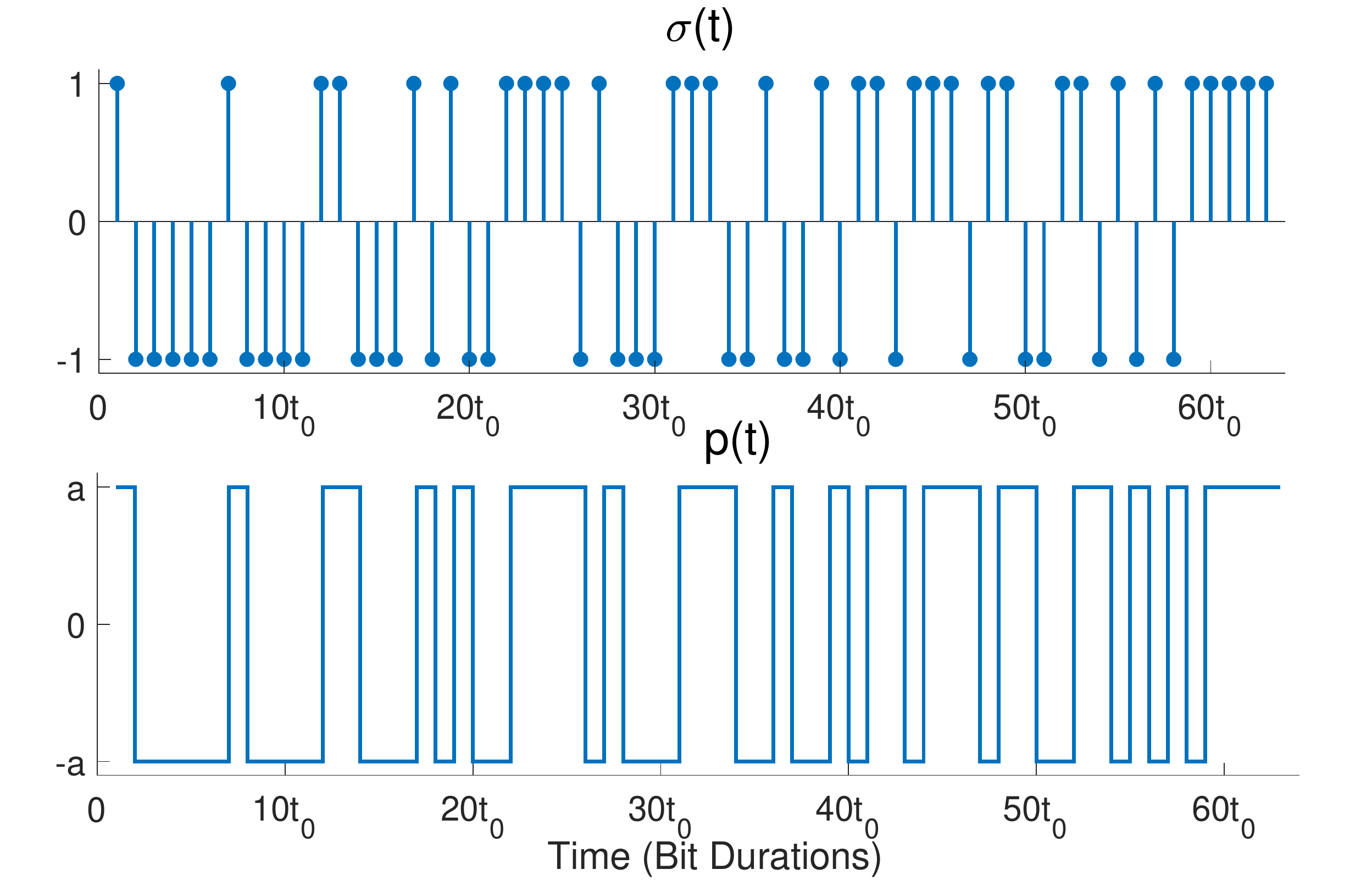}
    \caption{A single period of $\sigma(t)$, a pseudo-random binary impulsive sequence of order n = 6, and its corresponding probing signal, $p(t)$, a pseudo-random binary pulse train (PRBPT) \cite{NASA_Probe}. }
    \label{Probing_Block}
\end{figure}

In (\ref{probe_sig}), $\eta (t)$ is a rectangular pulse of unit magnitude and duration $t_0$, and $\alpha$ is a constant scaling factor.  $\sigma(t)$ is defined as a pseudo-random binary impulsive sequence, with order $n$, bit duration, $t_0$, and period $T_p = (2^n-1)*t_0$.

In addition to  the probing signal $p(t)$, the system is excited by its nominal input, $u(t)$. Assuming a linear time invariant system, the output can be then be expressed as $y(t) = y_{s}(t) + y_{p}(t)$, where $y_{s}(t) = h(t)*u(t)$ and $y_p(t) = h(t)*p(t)$.

A reference signal, $s(t)$, is taken to be a cyclic repetition of the probing signal $p(t)$, with magnitude $1/(\alpha T_p)$. For sufficiently small bit duration, $t_0$:

\begin{equation}
    p(t)\otimes s(t) \approx \sum_k  \delta(t-kT_p) 
    \label{xcorr}
\end{equation}

Here,  $\otimes$ represents cross-correlation and $\delta(t)$ is the Dirac delta function. The probing output is obtained by cross-correlating the reference signal with the system output: $z(t) = y(t) \otimes s(t) = z_{s}(t) + z_{p}(t)$, where by (\ref{xcorr}):

\begin{equation}
    z_{p}(t) = h(t)*p(t)\otimes s(t) \approx \sum_k  h(t-kT_p) 
    \label{xcorr2}
\end{equation}

In (\ref{xcorr2}), the probing output approximates a time-shifted repetition of the impulse response with period $T_{p}$. It is shown in \cite{NASA_Probe} that for sufficiently large $n$ or small $t_0$ (for a given $T_{p}$),  $z(t) \approx z_{p}(t)$ and for small $\alpha$, $y(t) \approx y_{s}(t)$.

In addition to these signal design requirements,  $t_{0}$ should be chosen to provide a sufficient sample rate to capture system dynamics, given the system's bandwidth $\omega_B$ as in (\ref{bit_durr}):

\begin{equation}
    t_0 \leq \frac{2\pi}{5\omega_B}
    \label{bit_durr}
\end{equation}

For a desired period, $T_p = (2^n-1)*t_0$, (\ref{bit_durr}) forces $n \geq log_2(T_p/t_0+1)$. $T_p$ should be chosen to be at least as long as the effective memory length of $h(t)$ to prevent aliasing.

The remaining probing signal parameter $\alpha$ is tuned along with $n$ and $t_{0}$ to provide a sufficiently large signal to noise ratios at the system and probing outputs, where the signal to noise ratio at the system output is defined as:

\begin{equation}
    SNR_y = \frac{rms\left[ y_s(t) \right] }{rms\left[ y_p(t) \right]}
    \label{SNR}
\end{equation}

\section{Application to Distribution Feeders}

This section develops a time domain probing signal to extract the impulse response of a distribution circuit, and provides initial simulation results from implementation on the IEEE 13 Node Distribution Test System \cite{13_node}.

\subsection{Probing Signal Design and Algorithm}

To probe a power system in time domain, the probing signal bit duration $t_0$ must be small enough and PRBPT order $n$ large enough that the 60 Hz phase voltage and currents will be averaged out by the cross correlation described in equation (\ref{xcorr}). The amplitude $\alpha$ must likewise be chosen small enough that the probing signal falls within typical noise level and does not interfere with the quality of delivered power. Bit duration $t_0$ must furthermore be slower than typical thyristor switching times such that the probing signal could be realized and implemented by a power electronic device.

Based on these constraints, the PRBPT used in this work was designed with order $n=10$, a probing amplitude $\alpha = 50 V$, whereas the nominal voltage for the feeder under test is 4160 Vrms phase-phase or 3387 Volts Peak phase to ground. The bit duration is selected at $t_0 = 100 \mu s$. This gives a period $T_p = 0.1023 s$, just over 6 cycles. As the effective memory length of the channel from the Phase A voltage to current was found to be roughly 50 $ms$, the probing signal is sufficiently long to extract the impulse response without aliasing.

\begin{figure} [h!]
    \centering
    \includegraphics[scale = 0.6]{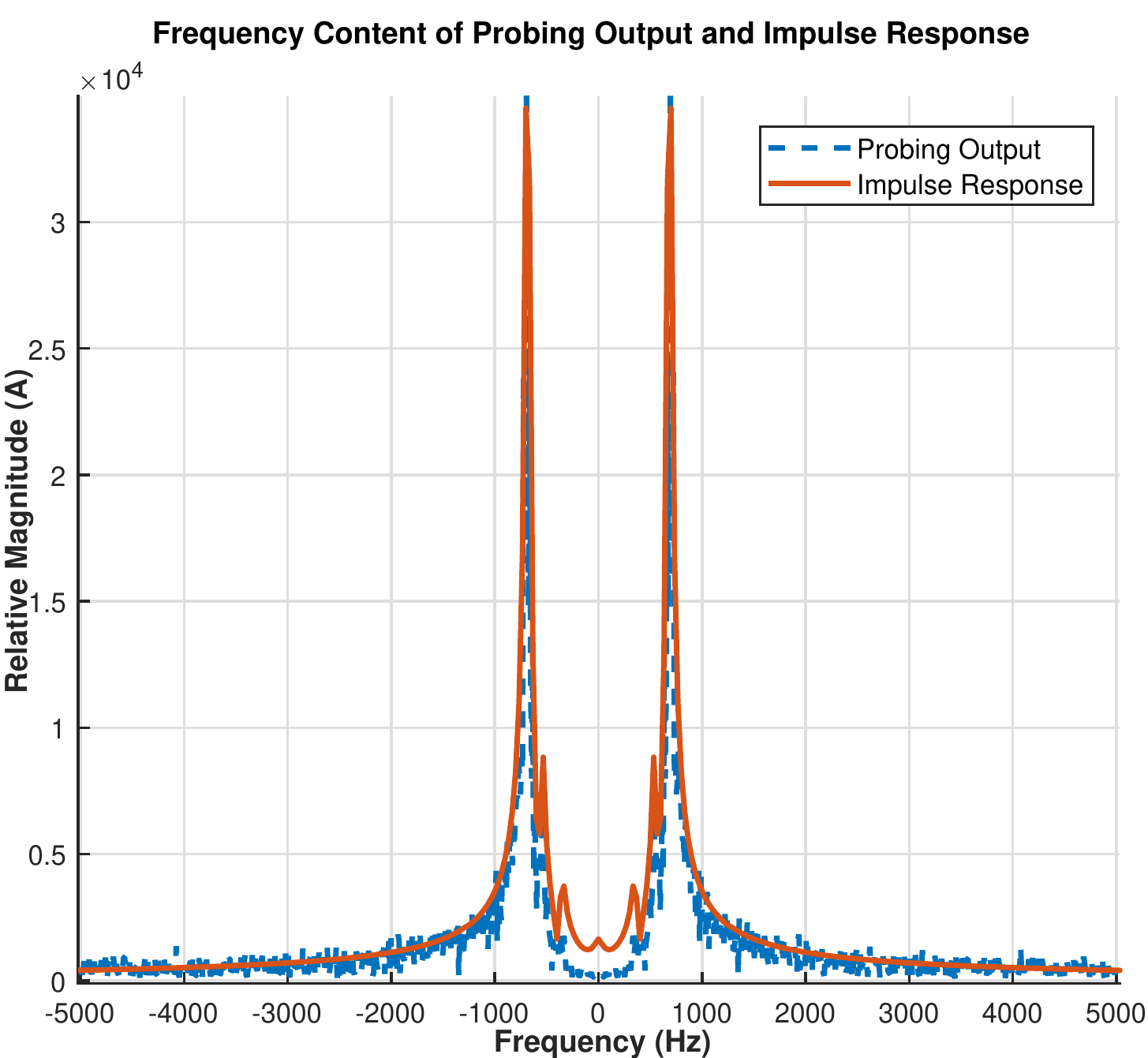}
    \caption{Fourier transform of the impulse response from Phase A voltage to Phase A current for the IEEE 13-Node Distribution Test Feeder \cite{13_node}. The impulse response obtained via applying a discrete time impulse to the system (red) is compared against the impulse response obtained from pulse compression probing (blue). Notably, with a notch filter for 60Hz, the probing algorithm is able to extract the impulse response during on-line operation.} 
    \label{Frequenct Content}
    \vspace{-3mm}
\end{figure}

Under this selection of parameters, the  60Hz component is still found in the probing output. This can be removed  for example, by raising $\alpha$ such that the response to the probing signal is larger, lowering $t_0$, or raising $n$ to average over more cycles. Each of these compromises an objective of the probing algorithm, which should inject a low-amplitude signal, switch on a timescale that could be feasibly implemented by thyristor switching, and ideally fast enough to be comparable to relay clearing times.

Rather than compromise one or more of these desirable features, a notch filter tuned to 60 Hz is applied to the system output $y(t)$ prior to the cross correlation in (\ref{xcorr}). Fig \ref{Frequenct Content} shows the frequency content of the Phase A voltage to current impulse response for the IEEE 13 Node Distribution Test Feeder obtained by direct test and by probing with an added notch filter. 

As PCP extracts the discrete time impulse response of a circuit with timestep $t_0$, the Hankel matrix can be constructed from the measured values at the probing output, which are exactly the Markov parameters of the system \cite{Linear_Book}:

\begin{equation}
    H = \begin{bmatrix}
    z_p(1) & z_p(2)& ... & z_p(m) \\
    z_p(2) & z_p(3)&... & z_p(m+1) \\
    \vdots  & \vdots & \ddots  & \vdots \\
    z_p(m) & z_p(m+1)  & ... & z_p(2m)
    \end{bmatrix} \label{Hankel}
\end{equation}
\normalsize

Here, $m$ is selected such that the impulse response has settled near zero before $2m$ time steps. With the Hankel Matrix expressed as in (\ref{Hankel}),  a state-space realization is readily available via a balanced realization or application of the Eigensystem Realization Algorithm \cite{ERA}.

The resulting discrete time transfer function is converted to continuous time. The steps of the probing process are summarized as follows:

  1. The probing signal is injected for a full period, $T_p$, and the system output is measured at the same sampling interval.
  
  2. The measured output $y(t)$ is fed through a notch filter tuned to 60Hz, then stored in memory.
  
  3. After the last bit is recorded, cross-correlation is performed between the recorded output $y(t)$ and the reference signal, $s(t)$, yielding the probing output $z_p(t)$.
  
  4. The Hankel Matrix (\ref{Hankel}) is constructed from  $z_p(t)$.
  
  5. The spectra of Hankel Singular Values, $H = U_{r}\Sigma_{r} V_{r}^T$ is used to find an appropriate order for a state space model. $U_{r}, V_{r}$, and $\Sigma_{r}$ are reduced to an appropriate order, $r$.
  
  6. A balanced realization is taken: $A = \Sigma_{r}^{-1/2}U_{r}^T\Tilde{H_{r}}V_{r}\Sigma^{-1/2}$, $B = \Sigma_{r}^{1/2}V_{r}$, $C = U_{r}\Sigma_{r}^{1/2}$, where $\Tilde{H}$ is the Hankel matrix starting from $z_p(2)$, and only the first element of B and C are taken for the single-input single-output probing system.

\subsection{Obtaining an Equivalent Circuit Model}

The transfer function of the IEEE 13-Node Distribution Test feeder from the perspective of the transmission system (as the ratio of the injected Phase A current into the feeder, I, to the Phase A voltage at the feeder terminal, V) can be closely fit using a second order transfer function, with a zero on the real axis and two complex poles, as given in (\ref{2ordtf}):

\begin{equation}
\frac{I}{V} = \frac{\alpha s + b}{s^2 + cs + d}
\label{2ordtf}
\end{equation}

\begin{figure} [h!]
    \centering
    \includegraphics[scale = 0.44]{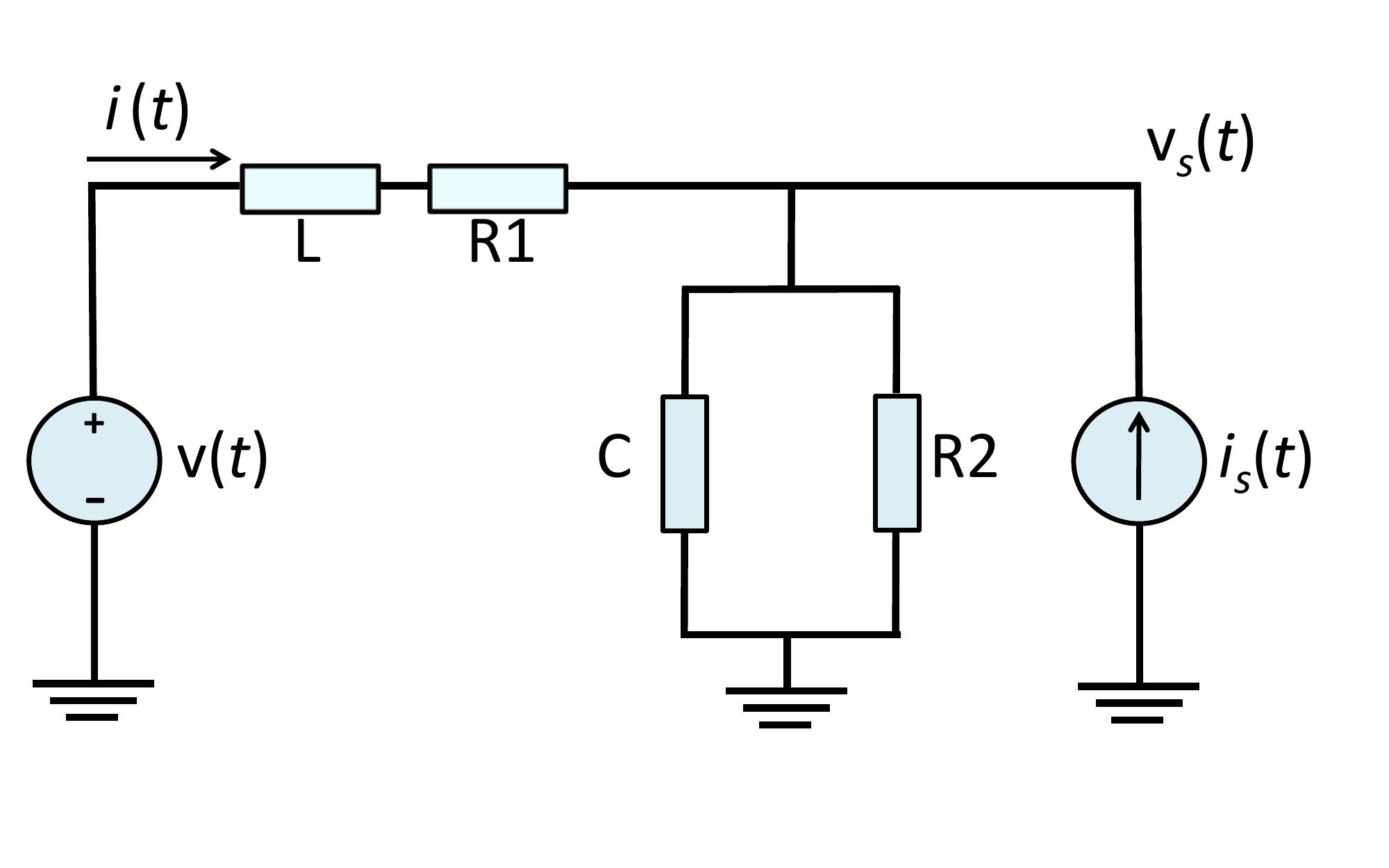}
    \vspace{-10mm}
    \caption{Equivalent electrical model of the distribution feeder with an aggregated distributed generation modeled as a current source, Is.} 
    \label{Circuit_fig}
    \vspace{-1mm}
\end{figure}

This transfer function can be implemented as an equivalent circuit using resistors, inductors, and capacitors. The circuit configuration used is shown in Fig. \ref{Circuit_fig}.
\vspace{-2mm}

The transfer function of the equivalent circuit shown in Fig. \ref{Circuit_fig} can be expressed as follows:

\begin{equation}
\frac{I}{V} = \frac{\frac{1}{L}s+\frac{1}{LR_2C}}{s^2 + \frac{L+R_1R_2C}{LR_2C}s + \frac{R_1+R_2}{LR_2C}}
\label{2ordtf_sub}
\end{equation}

By comparing (\ref{2ordtf}) and (\ref{2ordtf_sub}), it is clear that L is given by $L= 1/a,$ and the circuit’s remaining parameter values can be uniquely solved as:
\small
\begin{equation}
 R_1 = \frac{ac-b}{a^2}, R_2 = \frac{da^2-abc+b^2}{ba^2},\\ C = \frac{a^3}{da^2-abc+b^2}
\label{circuit params}
\end{equation}
\normalsize

In order to obtain the value of the equivalent source, $i_s (t)$ in Fig. \ref{Circuit_fig}, the concept of quasi–steady–state sinusoidal (QSSS) model is used \cite{QSSS}. The QSSS model tracks the states of an AC generating source, i.e. the source’s magnitude and angle. The QSSS model is based on the assumption that the magnitude and angle of a waveform are changing slowly with respect to the nominal frequency of the system (in North America the nominal frequency is 60 Hz). The addition of this QSSS source allows aggregate tracking of the real and reactive power output of loads and active sources within the feeder.

The QSSS modeled current source $i_s (t)$ can be represented as:
\begin{equation}
    i_s (t) = I_s (t) sin(\omega t+ \beta (t))
\label{QSSS}
\end{equation}

In (\ref{QSSS}) $\omega =2\pi f$, where f is the nomainal frequency of the system, and $I_s(t)$ and $\beta (t)$ are the magnitude and the angle of the current source, respectively. Similarly, one can obtain QSSS models of the feeder's terminal measurements, $v(t)$ and $i(t)$ as given in (\ref{time_to_phasor}).
\vspace{-1mm}
\begin{equation}
\begin{split}
        & v (t) = V(t) sin(\omega t + \theta(t))  \\
        & i (t)=  I(t) sin(\omega t + \alpha(t))
\end{split} \label{time_to_phasor}
\end{equation}
The unknowns, $V(t), \theta(t), I(t)$, and $\alpha(t)$ are estimated using time domain measurements and solving a set of linear equations. For the voltage, $V(t)$ and $\theta(t)$ can be solved for from initial time $\tau_0$ using (\ref{matrix_QSSS}).

\begin{equation}
\begin{bmatrix}
    v(\tau_0) \\
    v(\tau_0+\Delta t)\\
    \vdots\\
    v(\tau_0+n\Delta t)\\
\end{bmatrix} 
=
\begin{bmatrix}
    sin(\omega \tau_0)  cos(\omega \tau_0) \\
    sin(\omega \tau_0+\Delta t)  cos(\omega \tau_0+\Delta t)\\
    \vdots\\
    sin(\omega \tau_0+n\Delta t)  cos(\omega \tau_0+n\Delta t)\\
\end{bmatrix} 
\begin{bmatrix}
    A \\
    B\\
\end{bmatrix} 
    \label{matrix_QSSS}
\end{equation}

After solving for $A$ and $B$, $V(t)$ is computed as $V(t) =\sqrt{A^2+B^2}$  and $\theta (t) = atan(B/A)$.

To obtain $I_s(t)$ and $\beta (t)$, the circuit in Fig. \ref{Circuit_fig} is solved in phasor domain from terminal measurements $\hat{V} = V(t) \angle \theta (t)$ and $\hat{I} = I(t) \angle \alpha (t)$, as in (\ref{I_solution}):
\vspace{-1mm}
\begin{equation}
    \hat{I}_s = \frac{R_2C\omega+1}{R_2}\hat{V} - \left[ \frac{(R_2C\omega+1)(R_1+L\omega)}{R_2}+1  \right]\hat{I}
\label{I_solution}
\end{equation}
\section{Probing the IEEE 13 Node Test Feeder}

\begin{figure} [h]
    \centering
    \includegraphics[scale = 0.35]{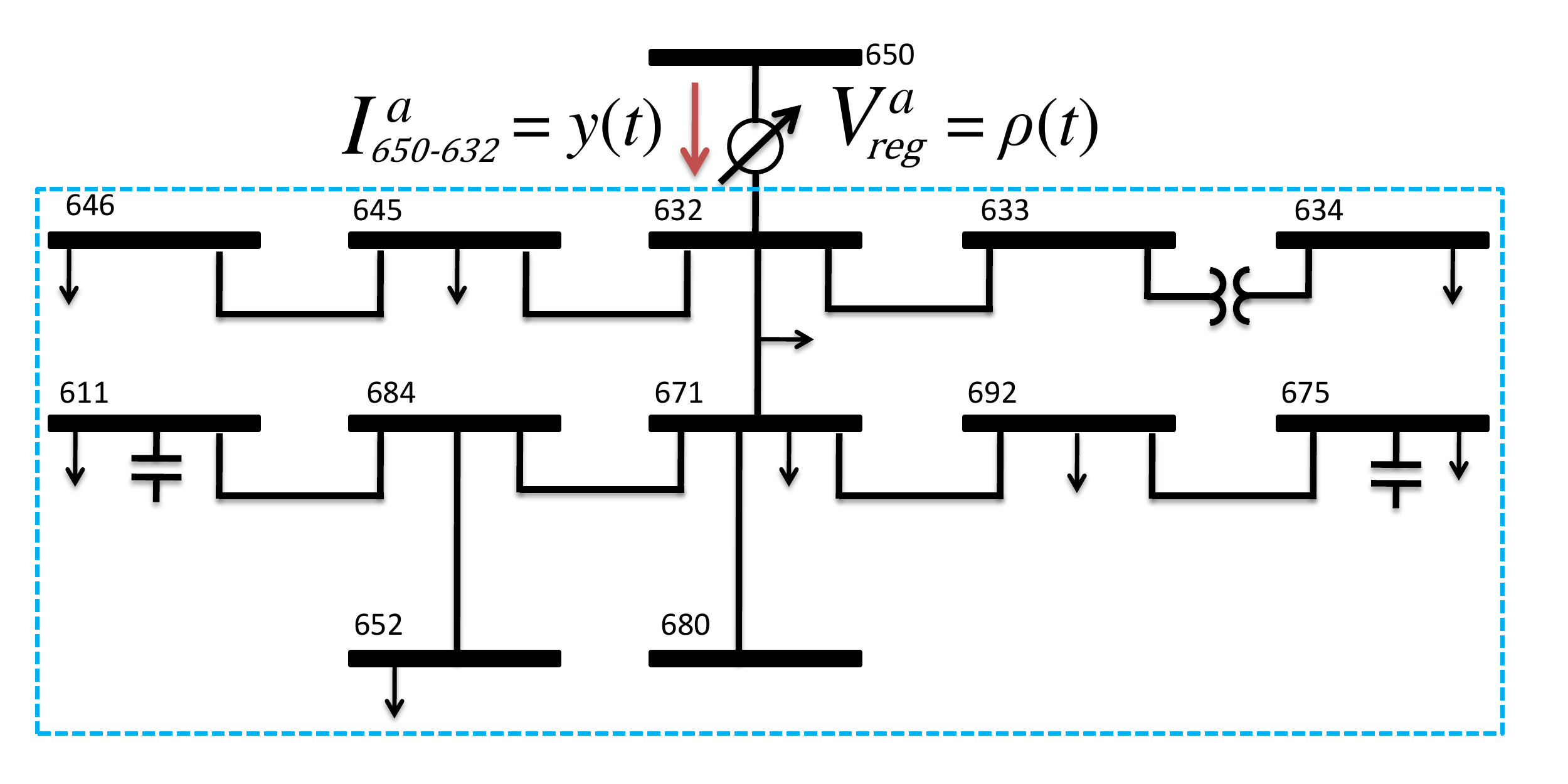}
    \vspace{-6mm}
    \caption{Single Line Diagram of the IEEE 13 Node Distribution Test Feeder \cite{13_node} with probing input and output identified. The blue border denotes the busses and lines captured by the probing and represented by the circuit model. The PRBPT input signal is injected as a voltage on Phase A over the voltage regulator between busses 650 and 632, and the Phase A current is taken as the output signal. The arrow between busses 632 and 671 identifies distributed loads.} 
    \label{13_node}
\end{figure}

To test the PCP algorithm, it is implemented on the IEEE 13 Node Distribution Test Feeder as shown in Fig. \ref{13_node}. 

From the figure, the probing signal is injected in series with Phase A of the voltage regulator between busses 650 and 632. As measurement devices are more likely to be available at the substation, the Phase A current entering the network is used as the output signal.

The probing signal source is added as a physical signal to the existing Simulink Model for the IEEE 13 Node Test Feeder \cite{Sybille}, and a measurement is added to the Phase A current between busses 650 and 632.

\begin{figure} [h!]
    \centering
    \includegraphics[scale = 0.55]{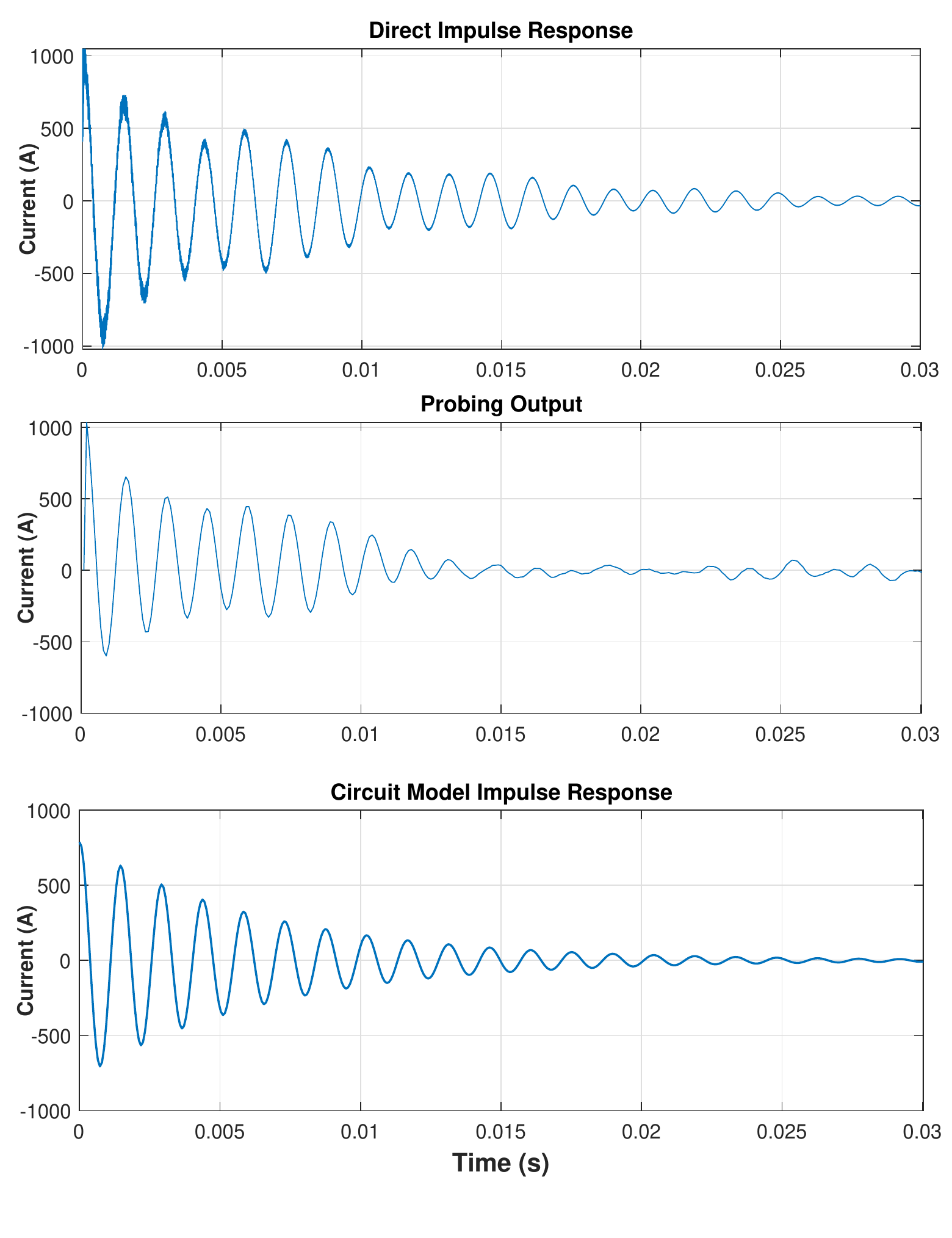}
  \vspace{-8mm}
    \caption{Comparison of voltage to current impulse response on Phase A for a direct test by applying a discrete time impulse, the probing output, and the 2nd order equivalent circuit model. Notably, measurement noise is evident at the tail of the probing output and both the probing output and circuit model exclude fast-decaying high frequency components.} 
    \label{Impulse_Response}
    \vspace{-2mm}
\end{figure}

Fig. \ref{Impulse_Response} shows a comparison of the impulse response recovered by using PCP with that observed via direct test, applying an impulse to the Phase A voltage. 
From the Figure, the probing response and circuit model miss the initial impulse, which is an artifact of the discrete time test, as well as some initial high frequency components.

It is clear that a first order model cannot represent this impulse response, which has a damped oscillation at 700 Hz. In practical applications, identifying the impulse response of the circuit is also useful for HVDC controls, which must be designed carefully to avoid exciting the network's resonant frequency when a step voltage change is caused.

Table 1 shows the parameter values for the second order ficticious circuit models derived for various conditions on the feeder network. Notably, $R_1$ is found negative in all cases, but the equivalent model is not representative of a physical circuit. The model is designed to allow a QSSS representation and to track feeder dynamics in a timely manner at the terminals of the transmission network, which is the intended target for future study. Voltages and currents within the model are ficticious.

 \begin{table}[h] 
    \centering
            \caption{Circuit Model Parameters for Several Cases}
    \begin{tabular}{||c|c|c|c|c||}
    \hline
        Case & L (mH) &$R_1 (\Omega)$  &$R_2 (\Omega)$  & C $(\mu F)$  \\
           \hline
                Normal Condition & 14.72&	-1.402&	24.58&	34.52\\
          \hline
          High-Z Source at Bus 680 & 14.72&	-1.402&	24.58&	34.52\\
          \hline
          Zero-Z Source at Bus 680 & 16.05&	-1.685&	26.15&	30.61\\
          \hline
           Outage of 671-692 & 31.17 &	-59.56 &	11380 &	0.04509\\
          \hline
          50\% load increase at 675 & 15.28 &	-1.575 &	24.7 &	32.37\\
          \hline
    \end{tabular}
     \vspace{0.1mm}
    \label{circuits}
    \vspace{-3mm}
\end{table}

From the table, adding a high impedance source at bus 680 had no impact on the impulse response of the feeder network or the parameters of the derived circuit model. This is an expected result in the sense that the PRBPT averages out slower signals in cross correlation (\ref{xcorr}), and a notch filter was applied. As such, voltage sources near 60 Hz are seen by the probing signal as a short circuit. 

The ability to detect system change in a space of four parameters, coupled with the short duration of the input PRBPT allows the distribution network model to be updated in near real time. From the perspective of the transmission network, this allows real time tracking of feeder dynamics due to load variation and DER intermittency.

\begin{figure} 
\vspace{-1mm}
    \centering
    \includegraphics[scale = 0.45]{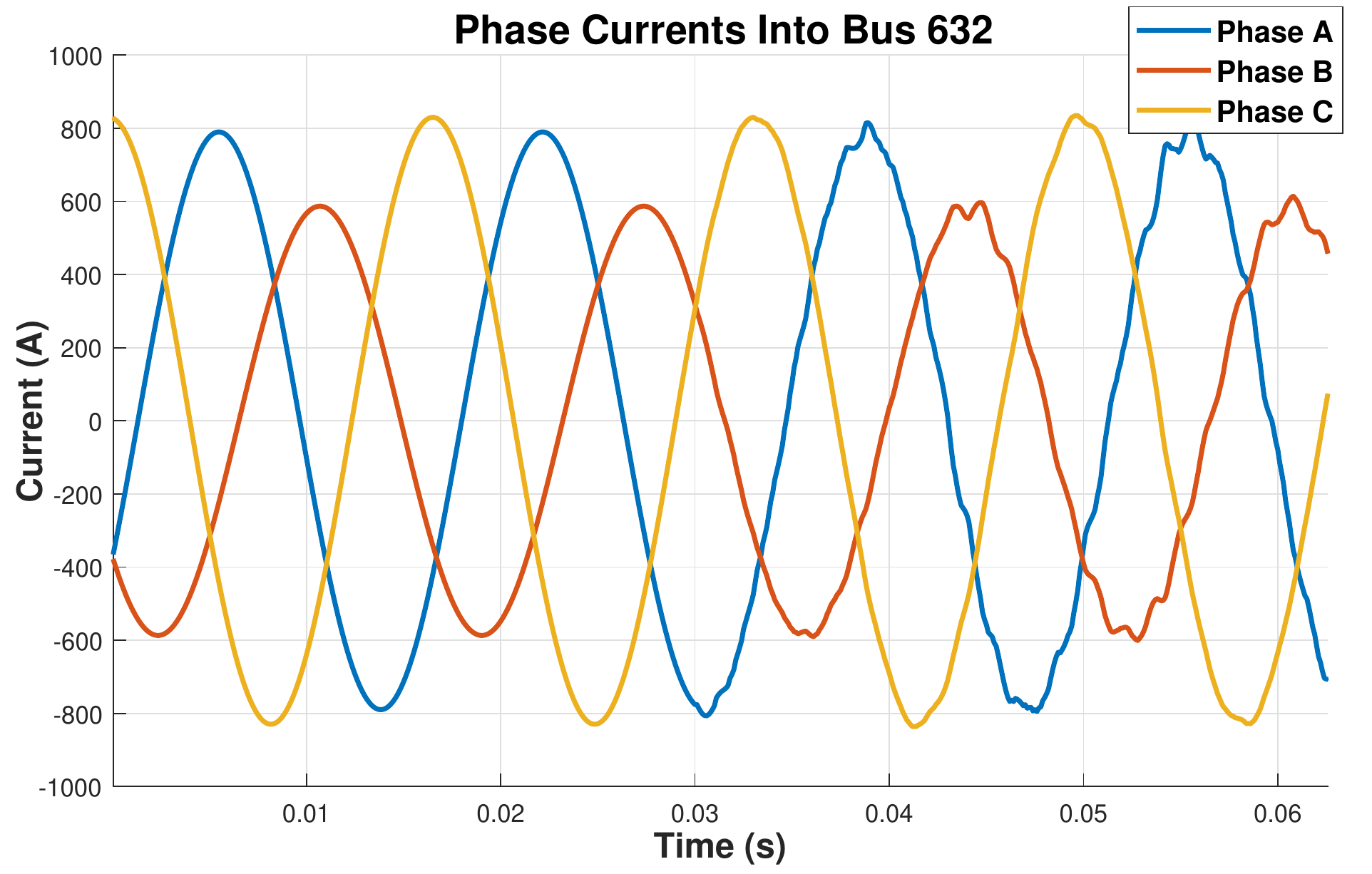}
    \caption{Phase currents flowing into Bus 632 during probing. The PRBPT signal is injected at time 0.03. The first two cycles show the phase currents without probing, and the third and fourth show the impact of the probing signal.} 
    \label{Current_SNR}
\end{figure}

The impact of the probing signal on the input current to the distribution feeder is shown in Fig. \ref{Current_SNR}. From the Figure, the impact of the PRBPT signal appears in the phase currents as a low level of noise. The Signal to Noise ratio in Phase A is computed using (\ref{SNR}) as 15.07 dB. By increasing the PRBPT order $n$, the magnitude of the probing signal, $\alpha$, can be reduced while maintaining the same probing output in (\ref{xcorr2}). This allows probing latency to be sacrificed in exchange for improved SNR.

\section{Discussion and Conclusion}
This work applies PCP to rapidly establish an equivalent model of a distribution feeder without requiring measurements within the feeder itself. For unbalanced networks such as the IEEE 13-bus test feeder used, an equivalent model can be established by probing each phase.

A pressing question for any power system probing algorithm is that of practical implementation. In particular, the fast bit duration makes the time-domain probing method presented in this paper require a particularly fast switching device. Flexible AC Transmission (FACTS) devices are identified as the best candidate to inject such a probing signal. In the particular case of the IEEE 13 Node Test Distribution system, a Thyristor-Controlled Voltage Regulator between busses 650 and 632, coupled with a current measurement device of sufficiently high sample rate, would be one possible means to implement the probing scheme. The detailed power electronics design to inject the PRBPT signal is left as future work.

Future work will involve on-line operation of the developed probing algorithm, and its use to frequently update equivalent distribution feeder models at the ports of a transmission network. On-line fault diagnosis in the transmission network, topology detection for distribution networks, as well as islanding detection are possible extensions using PCP.
\vspace{0mm}
\bibliography{bib}{}
\bibliographystyle{ieeetr}

\end{document}